\documentclass[11pt,a4]{article}
\usepackage{authblk}
\usepackage[numbers,sort&compress]{natbib}
\usepackage{notoccite}
\usepackage[nottoc]{tocbibind}
\usepackage{hyperref,setspace}
\usepackage{graphicx} 	
\usepackage{wrapfig}
\usepackage[usenames,table]{xcolor}

\date{}
\title{Snowmass White Paper: \\ Higher Spin Gravity and Higher Spin Symmetry}

\author{Xavier~Bekaert}
\author[2]{Nicolas~Boulanger}
\author[2]{Andrea~Campoleoni}
\author[3]{Marco~Chiodaroli}
\author[4]{Dario~Francia}
\author[5,6]{Maxim~Grigoriev}
\author[7]{Ergin~Sezgin}
\author[2,5]{Evgeny~Skvortsov}

\affil{Institut Denis Poisson, Unit\'e Mixte de Recherche $7013$ du CNRS,\protect\\ 
Universit\'e de Tours \& Universit\'e d'Orl\'eans,
Parc de Grandmount, 37200 Tours, France}

\affil[2]{Service de Physique de l'Univers, Champs et Gravitation,  
Universit\'e de Mons -- UMONS, \protect\\ 20 place du Parc, 7000 Mons, Belgium}

\affil[3]{Department of Physics and Astronomy,
Uppsala University, 75108 Uppsala, Sweden}

\affil[4]{Roma Tre University and INFN Roma Tre, via della Vasca Navale 84, 00146 Roma, Italy} 

\affil[5]{Lebedev Institute of Physics,
Leninsky ave. 53, 119991 Moscow, Russia}

\affil[6]{Institute for Theoretical and Mathematical Physics, \protect\\
  Lomonosov Moscow State University, 119991 Moscow, Russia  }

\affil[7]{Mitchell Institute for Fundamental Physics and Astronomy, \protect\\ Texas A\&M University, College Station, TX 77843, USA}

\begin{document}
\maketitle
\begin{abstract}
Higher Spin Gravity refers to extensions of gravity including at least one field of spin greater than two. These extensions are expected to provide manageable models of quantum gravity thanks to the infinite-dimensional (higher spin) gauge symmetry constraining them. One of the key aspects of Higher Spin Gravity/Symmetry is the range and diversity of topics it embraces: (a) higher spin fields play a role in quantum gravity, AdS/CFT, string theory and are expected to have important consequences in cosmology and black hole physics; (b) higher spin symmetry finds applications in Conformal Field Theories, condensed matter systems and dualities therein; (c)  these models often rely on tools developed in the study of the mathematical foundations of QFT or in pure mathematics: from deformation quantization and non-commutative geometry to conformal geometry, graded geometry (including BV-BRST quantization), and geometry of PDEs. Recent exciting applications also involve (d) modelling the coalescence of black hole binaries as scattering of massive higher spin particles.
\end{abstract}

\newpage
\tableofcontents
\newpage

\section{Introduction}

One of the major challenges of theoretical high energy physics has long been the  problem of quantum gravity. A number of approaches have emerged over the years, e.g.\
supergravities were born \begin{wrapfigure}{r}{0.30\textwidth}
    \includegraphics[width=0.3\textwidth]{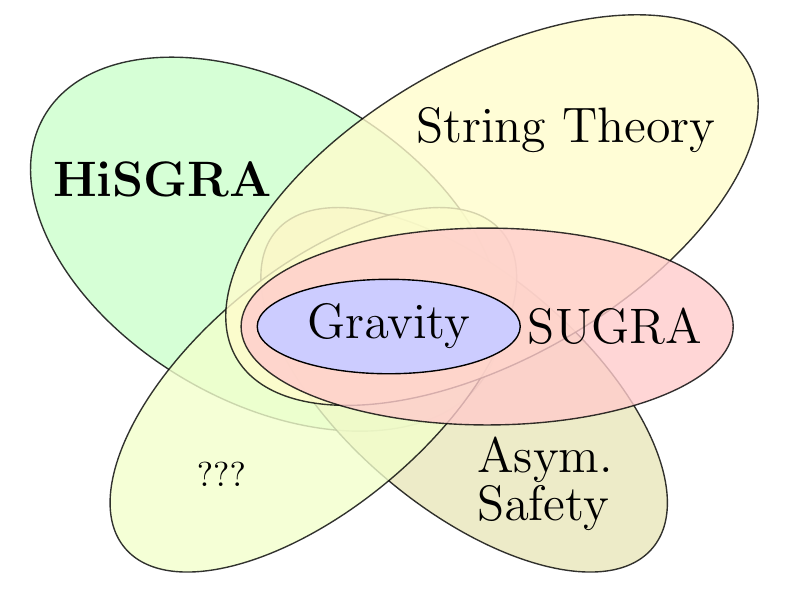}
\end{wrapfigure}
 with the idea that supersymmetry can improve the quantum behaviour 
of General Relativity \cite{VanNieuwenhuizen:1981ae},  which was followed by string theory, asymptotic safety
and few others, see e.g.~\cite{Nicolai:2013sz}. In the same vein as Supergravity, the founding idea
 of Higher Spin Gravity (HiSGRA) is to explore the most general gauge theories that can incorporate particles of any spin, hoping that the additional symmetries they bring in may improve their quantum behaviour \cite{Fradkin:1990kr}. Gauge theories, 
e.g.\ the Standard Model, and supergravities include massless particles with spin $s\leq2\,$, with $s=1\,$, $s=3/2$ and $s=2$ corresponding to gauge bosons, gravitini and
the graviton, respectively. 
On the other hand, the spectrum of String Theory has towers of (massive) higher spin states that are essential for its consistency. Independently of string theory, the AdS/CFT correspondence gives
strong indications that higher spin states (with $s > 2$) may be important to resolve the quantum gravity problem.
\begin{wrapfigure}{l}{0.32\textwidth}
    \includegraphics[width=0.32\textwidth]{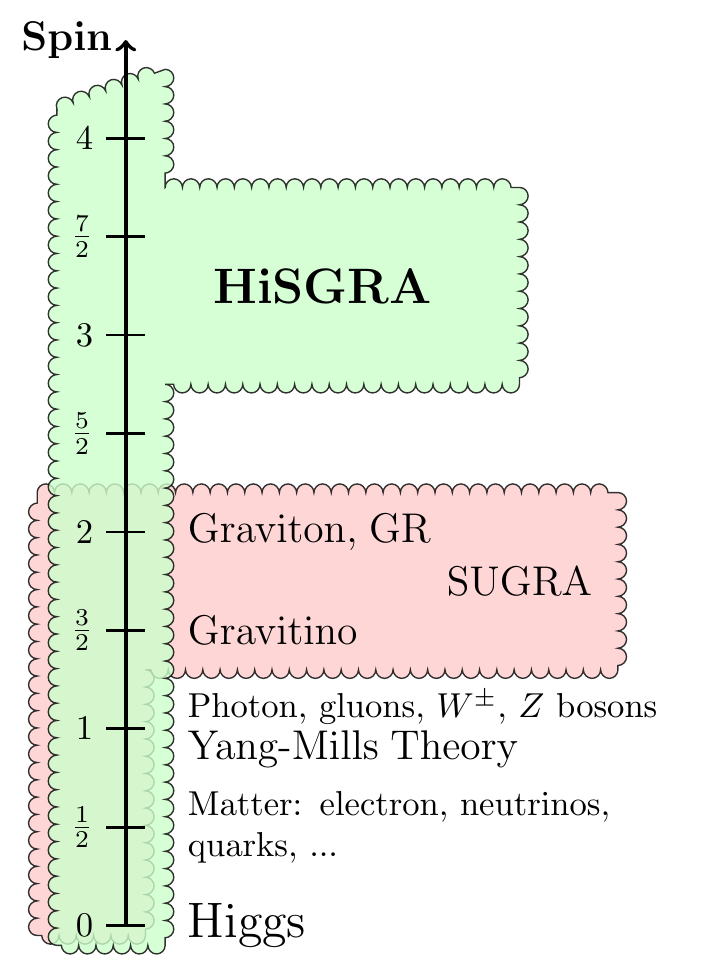}
\end{wrapfigure}

A remarkable and still not well-understood fact is that slightly different values 
of the spin lead to completely different theories and phenomena. They also imply different mathematical structures: gauge bosons of spin one are described by Yang-Mills theories and are, thus, parameterized by Lie groups and their representations; 
gauge bosons of spin two cannot have color   
and result in Einstein gravity plus higher derivative corrections \cite{Boulanger:2000rq}, 
which is governed by Riemannian geometry; 
gauge fermions of spin three-half lead to a zoo of supergravities 
\cite{VanNieuwenhuizen:1981ae}, which contain the graviton plus other fields and require 
superalgebras/supergeometry. 
The leitmotiv of HiSGRA is the first-principle systematical study of all possible fundamental 
interactions and their underlying symmetries and mathematical structures.

The presence of at least one massless higher spin field in the spectrum
implies that the theory must also contain the graviton, hence the name HiSGRA. 
In the simplest setups, the spectrum of HiSGRA actually contains massless particles 
of all integer spins, see e.g.~\cite{Fronsdal:1978rb,Flato:1978qz,Fradkin:1986ka,Konstein:1989ij,Vasiliev:1990en,Metsaev:1991mt,Metsaev:1991nb}. 
Masslessness severely constrains HiSGRA's 
since massless particles are described by gauge 
fields and interactions must respect the gauge symmetry. 
Interactions of massless higher spin particles are strongly constrained by the 
QFT principles.\footnote{The constraints are so strong, see e.g.~\cite{Maldacena:2011jn,Boulanger:2013zza,Alba:2015upa} for some modern results, 
that they are usually formulated as no-go theorems, e.g.~\cite{Weinberg:1964ew,Coleman:1967ad,Benincasa:2007xk,Porrati:2008rm,Bekaert:2010hp,Bekaert:2010hw,Fotopoulos:2010ay,Boulanger:2015ova,Roiban:2017iqg,Sleight:2017pcz,Ponomarev:2017nrr,Ponomarev:2017qab}, 
excluding whole classes of possible interacting field theories. } At the same time, 
masslessness can simulate some features of the UV regime,  
which makes HiSGRA's good probes of the Quantum Gravity Problem since the quantum issues 
are, to some extent, pushed to the classical level. 
 
Higher Spin Gravity has developed into a rich subject that has many 
links to other topics from fundamental physics to pure mathematics. 
Recent progress described below brought the subject to a new level of maturity, where concepts and methods are ready to be exported to other areas.

\section{Recent Progress and State of the Art}
Many of the future challenges are closely related to the {\bf Recent Progress}, which we summarize below. 

\begin{description}
    \item[Quantum corrections/UV finiteness.] HiSGRA has initially been studied 
    with the hope that the extended symmetries accompanying massless higher spin fields can cure the problems of General Relativity at the quantum level \cite{Fradkin:1990kr}. It is only recently that the computation of quantum corrections became feasible. Concrete results include the following: (a) Explicit calculations of vacuum one-loop corrections \cite{Gaberdiel:2011zw,Giombi:2013fka,Tseytlin:2013jya,Tseytlin:2013fca,Giombi:2014iua,Giombi:2014yra,Beccaria:2014jxa,Beccaria:2014xda,Beccaria:2014zma,Basile:2014wua,Beccaria:2014qea,Beccaria:2015vaa,Beccaria:2016tqy,Bae:2016rgm,Bae:2016hfy,Skvortsov:2017ldz}, the idea being to read off the spectrum of HiSGRA's from the expected holographic dual CFT's, e.g.\ various vector models \cite{Klebanov:2002ja,Leigh:2003gk,Sezgin:2003pt,Giombi:2011kc}. Knowing the spectrum one can compute the one-loop determinant of the kinetic operators. Higher spin multiplets are infinite,
    therefore a
    regularization is required in order to be able to sum up the individual one-loop contributions. It was found for a great number of cases that these (regularized) vacuum one-loop corrections either vanish or are proportional to the free energy of the CFT duals, which is consistent with AdS/CFT duality. These results reveal also that the infinite multiplets, as determined by higher spin symmetry, are crucial for the cancellation of divergences and anomalies. (b) AdS/CFT-inspired techniques combined with higher spin symmetry lead to a one-loop contribution to the four-point function that is consistent with (a) \cite{Ponomarev:2016jqk}. (c) One-loop amplitudes were computed in various simple models \cite{Ponomarev:2019ltz}, e.g.\ in Conformal HiSGRA \cite{Joung:2015eny,Beccaria:2016syk,Adamo:2018srx} and in Chiral HiSGRA \cite{Skvortsov:2018jea,Skvortsov:2020wtf,Skvortsov:2020gpn}, where there are encouraging results that higher spin symmetry improves the quantum behavior of amplitudes and can even render them finite.

    \item[Higher spin symmetry for CFT's.] It has been understood that, at least in the large-$N$ limit, a rich class of conformal field theories in three dimensions should be controlled by higher spin symmetry. The class of CFT's consists of (Chern-Simons) vector models and incorporates many important CFT's, e.g.\ the Ising model, which has been a major challenge for a number of decades. In the free or large-$N$ limit these conformal field theories are known to feature an infinite-dimensional extension of the conformal symmetry, which is exactly the higher spin symmetry \cite{Fradkin:1986ka,Eastwood:2002su}, as shown in \cite{Maldacena:2011jn,Boulanger:2013zza,Alba:2013yda,Alba:2015upa}.  This symmetry alone is powerful enough to fix all correlation functions  \cite{Maldacena:2011jn,Colombo:2012jx,Didenko:2012tv,Didenko:2013bj,Bonezzi:2017vha}. A challenge has been to understand what happens when interactions are turned on or when one departs from the large-$N$ limit. Due to the very sparse spectrum of operators in vector models, the conserved tensors (higher spin currents) that are responsible for the higher spin symmetry can have few terms that violate the conservation. Importantly, the non-conservation operator is a composite operator built from the higher spin currents themselves. This allows one to close the loop and to bootstrap the correlation functions from the symmetry vantage point, putting aside any concrete details of the microscopical realization. This idea is called slightly-broken higher spin symmetry \cite{Maldacena:2012sf}. It seems to be powerful enough to fix three-point and four-point correlators  \cite{Maldacena:2012sf,Giombi:2016zwa,Turiaci:2018nua,Skvortsov:2018uru,Li:2019twz,Kalloor:2019xjb,Silva:2021ece,Jain:2021gwa}. It has also been important to understand what slightly-broken higher spin symmetry means algebraically and the proposal is that it is realized as a certain strong homotopy algebra \cite{Sharapov:2018kjz,Gerasimenko:2021sxj}. 
    
    \item[Tensionless limit of string theory. ] Since the early days of HiSGRA there has been a strong indication that certain HiSGRA's can emerge in the tensionless limit of string theory or, to put it in reverse, that string theory could be understood as a spontaneously broken phase of some HiSGRA \cite{Gross:1987kza,Gross:1987ar,Bonelli:2003kh,Sagnotti:2003qa,Sagnotti:2011jdy}.  Also, phenomenological applications bring to the forefront the important question of how higher spin symmetry breaking can occur. Whereas the systematics of such breaking and its possible physical implications for string theory and beyond are still among the main open questions in the subject, some concrete proposals on how this can be realized have started to emerge thanks to the AdS/CFT correspondence \cite{Sundborg:2000wp,Mikhailov:2002bp,Sezgin:2002rt,Bonelli:2003zu,Gaberdiel:2014cha,Gaberdiel:2015wpo}.

    The appearance of massless higher spin states in the spectra of strings and branes in four and more spacetime dimensions was studied using semi-classical methods in \cite{Gubser:2002tv,Sezgin:2002rt}, and using discretization methods in  \cite{Engquist:2005yt}, leading to dual formulations in terms of non-compact Wess--Zumino--Witten models with critical levels \cite{Engquist:2007pr}, that are solvable by means of spectral flows and admit handy free-field realizations in terms of symplectic bosons with W-gauge symmetries. 
    The holography proposals have since been sharpened in \cite{Chang:2012kt} in the case of $AdS_4$, and the tensionless string proposal has recently been made more concrete in $AdS_5$ in  \cite{Gaberdiel:2021qbb}. In $AdS_3$, limits involving tensionless strings with NS fluxes have led to concrete examples of complete AdS/CFT correspondences \cite{Gaberdiel:2018rqv}, including entire string spectra and not just truncations to massless states or first Regge trajectories.

    \item[Higher spin models in low dimensions.]
    
    In three spacetime dimensions the little group of massless particles does not admit non-trivial irreducible representations of arbitrary helicity. Nevertheless, one can consider, e.g., the couplings of symmetric tensor fields to gravity or between themselves, and build low-dimensional models for HiSGRA \cite{Aragone:1983sz}. The simplest examples generalise the Chern-Simons formulation of three-dimensional gravity and do not require an infinite number of higher spin fields \cite{Blencowe:1988gj,Bergshoeff:1989ns}. This remarkable simplification allowed to show that the asymptotic symmetries are given by ${\cal W}$-algebras \cite{Henneaux:2010xg,Campoleoni:2010zq} and to build solutions carrying entropy, identified as higher spin black holes \cite{Gutperle:2011kf,Ammon:2011nk,Gaberdiel:2012yb,Perez:2012cf,Campoleoni:2012hp,deBoer:2013gz,Bunster:2014mua,Grumiller:2016kcp}. Customary signatures of black holes, like metric singularities and event horizons, are, however, not invariant under higher spin gauge transformations: these solutions thus provide a natural playground to study problems like singularity resolution in extension of (quantum) gravity \cite{Ammon:2011nk,Castro:2011fm}. The previous results were first established on AdS$_3$ and accompanied by generalisations of various holographic tools, see e.g.~\cite{Ammon:2013hba,deBoer:2013vca,Perlmutter:2013paa,deBoer:2014sna,Perlmutter:2016pkf,Narayan:2019ove,Alday:2020qkm,Zhao:2022wnp}. On the other hand, in three dimensions one can also build along the same lines flat-space HiSGRAs \cite{Blencowe:1988gj,Campoleoni:2011tn,Afshar:2013vka,Gonzalez:2013oaa}. Similar generalisations exist for conformal gravity as well \cite{Pope:1989vj,Fradkin:1989xt,Grigoriev:2019xmp}.

    The appearance of ${\cal W}$-symmetries led to conjecture a holographic duality between higher spin gauge theories coupled to matter \cite{Prokushkin:1998bq} and ${\cal W}_N$ minimal models \cite{Gaberdiel:2010pz}. Given the good degree of control over ${\cal W}_N$ minimal models that have been accumulated over the years, the duality has been subjected to several tests \cite{Gaberdiel:2011wb,Gaberdiel:2011zw,Kraus:2011ds,Creutzig:2011fe,Ammon:2011ua,Gaberdiel:2012ku,Fredenhagen:2018guf}, that confirmed its bases and allowed further subtle refinements to account for quantum effects \cite{Chang:2011mz,Castro:2011iw,Perlmutter:2012ds}. The key issue which remains to be understood is whether minimal model holography can stand on its own even at the quantum level or if it can only be considered as an effective duality emerging in the tensionless limit of suitable compactifications of string theory on AdS$_3$ \cite{Gaberdiel:2014cha}. For this reason, minimal model holography also sparkled an in-depth analysis of the tensionless limit of string theory on AdS$_3$, see e.g.~\cite{Gaberdiel:2013vva,Gaberdiel:2018rqv,Eberhardt:2018ouy}.
    
    Interestingly, in two dimensions, there exists a similar higher spin extension of Jackiw-Teitelboim gravity \cite{Alkalaev:2013fsa,Grumiller:2013swa,Alkalaev:2014qpa,Alkalaev:2020kut} exploiting the fact that the latter is a BF theory. Two-dimensional higher spin gravity is an almost uncharted territory which is worth exploring along similar lines to three-dimensional one, especially due to its potential relevance for building a bulk dual of the Sachdev–Ye–Kitaev model \cite{Gonzalez:2018enk,Alkalaev:2019xuv,Datta:2021efl,Peng:2018zap,Kruthoff:2022voq}.
    
    \item[Massive higher spin fields for Black Hole scattering. ]   Very recent excitement about higher spin theories is due to gravitational wave physics, where massive higher spin particles can be used to model rotating black holes \cite{Arkani-Hamed:2017jhn,Guevara:2018wpp,Maybee:2019jus,Guevara:2019fsj,Arkani-Hamed:2019ymq,Bern:2020buy,Chiodaroli:2021eug}. The recent experimental discovery of gravitational waves by LIGO and the rapid development of gravitational wave physics leads to new challenges for theoretical physics. There is a high demand for efficient methods to compute various characteristics of complicated processes in General Relativity to many orders in post-Minkowskian and post-Newtonian expansions. These challenging and timely computations can be done efficiently provided a HiSGRA in four-dimensional spacetime is available whose spectrum contains massive higher spin fields and the graviton.

\end{description}

\noindent It is also worth giving a concise list of the results achieved within the topic so far as to summarize the current {\bf State of the Art}:
\begin{itemize}
\item  Many different approaches to higher spin fields 
have been developed
over the years: metric-like \cite{Singh:1974qz,Singh:1974rc,Fronsdal:1978rb,Fang:1978wz,deWit:1979sib}, generalizing the metric as a tool to describe a massless spin-two field to any spin;  
frame-like \cite{Aragone:1979hx,Aragone:1980rk,Vasiliev:1980as,Fradkin:1986qy,Lopatin:1987hz,Zinoviev:2010cr,Boulanger:2012dx}, extending tetrad/vielbein and spin-connection variables to any spin;  light-front \cite{Bengtsson:1983pd,Bengtsson:1983pg,Bengtsson:1986kh,Metsaev:1991mt,Metsaev:2005ar,Metsaev:1991nb,Metsaev:2018xip}; unfolded 
\cite{Vasiliev:1988sa,Vasiliev:1990vu,Bekaert:2004qos}, based on Free Differential Algebra techniques \cite{Sullivan77,DAuria:1980cmy,Nieuwenhuizen:1982zf};
parent formalism~\cite{Barnich:2004cr,Barnich:2006pc,Bekaert:2009fg,Alkalaev:2009vm,Barnich:2010sw,Grigoriev:2010ic,Alkalaev:2013hta}, unifying unfolded and metric-like approaches within the unified Batalin--Vilkoviski--Becchi--Rouet--Stora--Tyutin (BV-BRST) scheme that can also be seen as a generalization of the Alexandrov--Kontsevich--Schwarz--Zaboronsky (AKSZ) approach~\cite{Alexandrov:1995kv}; Maxwell-like \cite{Francia:2010qp,Campoleoni:2012th,Francia:2013sca,Bekaert:2015fwa,Francia:2016weg}; 
twistor and twistor-inspired \cite{Penrose:1965am,Hughston:1979tq,Eastwood:1981jy,Woodhouse:1985id,Krasnov:2021nsq}, which are suitable for self-dual higher spin theories.

\item  
One of the most powerful approaches to constructing 
field theories bottom-up that has been developed is the perturbative Noether procedure 
\cite{Berends:1984wp,Berends:1984rq} that 
received a systematic BV-BRST reformulation in \cite{Barnich:1993vg}, see also \cite{Barnich:1995db,Barnich:2000zw}.

\item  A detailed description of cubic (and some higher) interactions, which should be understood as 
seeds of interactions from which complete theories can grow, has been obtained within 
various approaches \cite{Bengtsson:1983pd,Bengtsson:1983pg,Berends:1984wp,Berends:1984rq,Bengtsson:1986kh,Fradkin:1987ks,Metsaev:1991mt,Alkalaev:2002rq,Bekaert:2005jf,Metsaev:2005ar,Boulanger:2006gr,Francia:2007qt,Metsaev:2007rn,Fotopoulos:2008ka, Zinoviev:2008ck,Boulanger:2008tg, Manvelyan:2010wp,Sagnotti:2010at,Fotopoulos:2010ay,Manvelyan:2010jr,Manvelyan:2010je,Joung:2011ww,Metsaev:2012uy,Joung:2013nma,Bekaert:2014cea,Bekaert:2015tva,Sleight:2016dba,Conde:2016izb,Francia:2016weg,Buchbinder:2017nuc,Mkrtchyan:2017ixk,Buchbinder:2018wzq,Metsaev:2018xip,Joung:2019wbl,Fredenhagen:2019hvb,Metsaev:2019dqt,Metsaev:2019aig,Fredenhagen:2019lsz,Grigoriev:2020lzu,Buchbinder:2021xbk}. The complete classification of cubic interactions involving massless as well as massive, fermionic and bosonic, fields of any spin is available.

\item  Extensions of spacetime symmetries, known as higher spin symmetries, 
have been thoroughly studied \cite{Dirac:1963ta, Gunaydin:1981yq,Gunaydin:1983yj,Fradkin:1986ka,Vasiliev:1986qx, Gunaydin:1989um,Nikitin1991,Konstein:2000bi,Eastwood:2002su} and were related to deformation quantization \cite{Iazeolla:2008ix,Boulanger:2011se,Boulanger:2013zza,Michel,Joung:2014qya,Joung:2015jza} and to higher symmetries of PDE's \cite{Nikitin1991,Eastwood:2002su, Grigoriev:2006tt,Bekaert:2009fg,Michel}. It was also proved that rigid higher spin symmetry is realized in free conformal field theories \cite{Maldacena:2011jn,Boulanger:2013zza,Alba:2013yda,Alba:2015upa}. In particular, the correlation functions can be represented and computed as higher spin invariants \cite{Colombo:2012jx,Didenko:2012tv,Didenko:2013bj,Bonezzi:2017vha}.  Recently, higher spin asymptotic symmetries, providing counterparts of the infinite-dimensional symmetries emerging for spin-one gauge systems and for gravity in asymptotically flat spaces, have been identified at the linear level in \cite{Campoleoni:2017mbt, Campoleoni:2020ejn} and a candidate non-Abelian algebra for their global part was proposed \cite{Campoleoni:2021blr}. 
\item  Higher spin gravities found a natural place within the AdS/CFT paradigm as 
gravitational duals of simple conformal field theories 
\cite{Bergshoeff:1988jm,Sundborg:2000wp,Mikhailov:2002bp, Sezgin:2002rt,Klebanov:2002ja,Sezgin:2003pt,Leigh:2003gk,Giombi:2011kc}, 
including the free ones. One large class of theories is given by vector models, e.g.\ by the free and critical vector models. Another class incorporates tensionless limits of string theories and M-theory \cite{Bergshoeff:1988jm,Sundborg:2000wp, Sezgin:2002rt, Chang:2012kt}.

\item The latter HiSGRA AdS/CFT developments \cite{Giombi:2009wh,Giombi:2010vg,Giombi:2011ya} triggered the discovery of the three-dimensional bosonization duality \cite{Giombi:2011kc, Maldacena:2012sf, Aharony:2012nh,Aharony:2015mjs,Karch:2016sxi,Seiberg:2016gmd}. More generally, a large class 
of $3D$ conformal field theories, known as Chern-Simons vector models, are related by a number of dualities. These models exhibit in 
the large-$N$ limit a peculiar realization of the higher spin symmetry as an 
infinite-dimensional extension of the conformal one, called slightly-broken higher spin symmetry \cite{Maldacena:2012sf}. This new type of a symmetry seems to be powerful enough to fix correlation functions at least in the large-$N$ limit and, thereby, explain the three-dimensional bosonization duality \cite{Giombi:2016zwa,Turiaci:2018nua,Li:2019twz,Kalloor:2019xjb,Jain:2021gwa,Silva:2021ece}. 

\item  Within the formal deformation approach many higher spin gravity models have been constructed \cite{Vasiliev:1990en,Vasiliev:2003ev,Sagnotti:2005ns,Boulanger:2011dd,Bekaert:2013zya,Brust:2016zns,Bonezzi:2016ttk,Bekaert:2017bpy,Grigoriev:2018wrx,Sharapov:2019pdu}, the problem was solved in full generality --- it was shown how to construct formally consistent classical equations for any given higher spin algebra  \cite{Sharapov:2019vyd}  --- and turned out to be deeply related to deformation 
quantization, topological field theory and non-commutative geometry \cite{Engquist:2005yt,Boulanger:2011dd,Sezgin:2011hq,Boulanger:2015kfa,Bonezzi:2016ttk,Li:2018rnc,Sharapov:2019vyd}. 

\item  Within the latter paradigm some exact solutions, including the domain wall, 
FLRW-like and the black-hole$\,$-like 
were found 
\cite{Sezgin:2005hf,Sezgin:2005pv,Didenko:2009td,Iazeolla:2012nf,Iazeolla:2017dxc,Aros:2017ror,Aros:2019pgj,DeFilippi:2021xon}.
The black-hole like solutions look promising in resolving the black-hole singularity problem. 

\item Concrete simple models of HiSGRA have been constructed, including Chern-Simons \cite{Blencowe:1988gj,Bergshoeff:1989ns,Fradkin:1989xt,Pope:1989vj,Henneaux:2010xg,Campoleoni:2010zq,Grigoriev:2019xmp,Grigoriev:2020lzu}, conformal \cite{Segal:2002gd,Tseytlin:2002gz,Bekaert:2010ky}, chiral \cite{Metsaev:1991mt,Metsaev:1991nb,Ponomarev:2016lrm,Skvortsov:2018jea,Skvortsov:2020wtf} HiSGRA's,
and their quantum corrections were found to reveal remarkable cancellations of UV divergences  \cite{Giombi:2013fka,Giombi:2014iua,Giombi:2014yra,Beccaria:2014xda,Beccaria:2014jxa,Beccaria:2014zma,Basile:2014wua,Beccaria:2014qea,Beccaria:2015vaa,Beccaria:2016tqy,Bae:2016rgm,Bae:2016hfy,Skvortsov:2017ldz,Skvortsov:2018jea,Skvortsov:2020wtf,Skvortsov:2020gpn,Ponomarev:2019ltz}. Other interesting recent proposals include \cite{Sperling:2017dts,deMelloKoch:2018ivk,Aharony:2020omh}.

\item  Some HiSGRA models were studied in regard to cosmological applications \cite{Anninos:2013rza,Aros:2017ror,Anninos:2019nib,Kim:2019wjo}. 

\item In addition to the study of massless fields, there is a number of systematic results on massive higher spin fields including cubic interactions \cite{Singh:1974qz,Singh:1974rc,Buchbinder:1999ar,Buchbinder:2000fy,Zinoviev:2001dt,Buchbinder:2005ua,Metsaev:2005ar,Buchbinder:2006ge,Zinoviev:2006im,Metsaev:2007rn,Zinoviev:2009gh,Ponomarev:2010st,Porrati:2010hm,Zinoviev:2011fv,Metsaev:2012uy,Buchbinder:2012xa,Zinoviev:2008ck,Conde:2016izb,Buchbinder:2017izy,Buchbinder:2019dof}, which are most relevant for applications of massive higher spin fields to black hole scattering.

\end{itemize}

\section{Future Directions}

\subsection{Higher Spin Gravity and Quantum Gravity}

One of the main future directions is to fulfil the original idea behind HiSGRA --- to prove that higher spin symmetry is powerful enough to make higher spin gravities renormalizable, or even finite, models of quantum gravity. To this end, the already existing results need to be pushed to higher loop orders, which presents a considerable technical challenge. Therefore, indirect, symmetry-based approaches can turn out to be more efficient.  For example, it should be possible to classify all possible higher spin invariant counterterms by computing the corresponding Chevalley-Eilenberg cohomology of higher spin algebras. The main hope would be that the corresponding groups are empty (the theory is finite) or contain just one representative corresponding to the on-shell action (the theory is renormalizable). In the same vein, it should be possible to approach possible anomalies. 

It is also important to enlarge the family of HiSGRA's with new models. A completely new fruitful direction within the HiSGRA approach would be to extend the existing results and techniques to a broader class of theories that are somehow in between the HiSGRA’s and string theory, i.e., to systematically look for theories that are much simpler than string theory, are closer to HiSGRA, but whose spectrum contains massive higher spin states. This study should clarify to which extent string theory can be regarded as the unique solution to the quantum gravity problem under some 
natural assumptions \cite{Caron-Huot:2016icg}. A particular class of such theories --- four-dimensional theories with one graviton and massive higher spins in the spectrum --- can model various processes in General Relativity, including scattering of black holes; see section \ref{bh}.

In addition, complete and simple enough HiSGRA’s models of quantum gravity should help to resolve the puzzles of cosmology of the early Universe. The main experimental challenge would be to look for deviations from Einstein gravity during the early stages of the Universe, while HiSGRA make specific predictions for higher derivative corrections to Einstein-Hilbert action with certain cosmological implications \cite{Anninos:2019nib}. The very notion of black holes has to be reconsidered in quantum gravity. HiSGRA's provide a concrete testing ground for ideas on quantum black holes and singularity resolution, and may help to shine new light on some of the old paradoxes of black hole physics, which, to some extent, has already been realized in $3D$ HiSGRA and needs to be extended to higher dimensions.

\subsection{Supersymmetry, Higher Dimensions and M-theory}      

At the fundamental level, it is important to understand whether and how higher spin symmetries could cure the problems of Einstein's gravity at the quantum level. In that context,  it is natural  to focus on pure HiSGRA  as the fundamental extension of  Einstein's gravity. At the same time, it is also  important to investigate the spontaneous breaking of higher spin symmetry~\cite{Girardello:2002pp} down to  the usual diffeomorphisms and gauge symmetries, and search for a unified description of matter couplings. 

While string theory with desirable properties requires supersymmetry and ten dimensions, so far no consistency considerations have been suggested to determine if similar constraints arise in HiSGRA's. Whether such constraints exist remains to be investigated. Besides consistency, the question of 
whether higher spin symmetry is powerful enough to render quantum gravity finite, or renormalizable, will need to be settled.  Concomitant to this, the question of how HiSGRA's  constructed so far can be constrained in search for a realistic (super) HiSGRA theory will require extensive studies.
These are among the key  questions  which deserve in depth studies and which will  also sharpen the queries on whether HiSGRA is  necessarily a phase of string/M-theory. In particular whether HiSGRA can be viewed as the tensionless limit of string theory and how  turning  on the string tension breaks the 
higher spin symmetry \cite{Sundborg:2000wp,Sezgin:2002rt,Mikhailov:2002bp,Bonelli:2003zu,Bonelli:2003kh,Girardello:2002pp,Sagnotti:2011jdy,Bianchi:2003wx,Beisert:2004di} will be among the important topics to explore. 

One of the most promising approaches for the description of HiSGRA's,  breaking of higher spin symmetry, and possible connections with string/M theory, is furnished by higher spin holography.  In particular, super ABJ models~\cite{Aharony:2008gk} in suitable limits have led to studies of slightly-broken higher spin symmetry~\cite{Hirano:2015yha,Binder:2021cif}, a (nonlocal) action description of HiSGRA in the bulk~\cite{Aharony:2020omh,deMelloKoch:2018ivk},  and possible connections with string/M theory compactifications. This is a fertile area which awaits several advances.

One tantalizing prospect for a connection with M-theory may also come from the exploration of relationship between the super HiSGRA 
based on higher spin extension of ${\cal N}=8$ super AdS group in $4D$, and the tensionless limit of supermembrane  in M-theory compactified on 
$AdS_4 \times S^7$. Certain aspects of this connection have been considered in \cite{Bergshoeff:1988jm,Sezgin:2002rt}  but much remains to be 
investigated.  
Whether the remarkable connection found between $3D$ SCFT and $11D$ supergravity correlators~\cite{Alday:2021ymb} can play a role in the study of these higher spin theories remains to be seen.

Regarding  the introduction of Yang-Mills gauge symmetries in HiSGRA's,  the key ingredients are singletons living on the boundary of AdS that carry representations of a suitable gauge group, and the holographic  constructions of bulk HiSGRA's \cite{Konstein:1989ij}.  Several details and extensions are yet to be worked out. As for the coupling of massive higher spin multiplets, their kinematics is based on the triple and higher order products of the singleton representations.  
A construction has been given 
in \cite{Vasiliev:2018zer}, see also \cite{Engquist:2005yt,Engquist:2007pr}, for their couplings by making use of multi-singleton oscillators, and connection to string theory has been conjectured.  Another approach may be the construction of fully nonlinear higher dimensional (super) HiSGRA's, and study of their compactifications. Fully nonlinear bosonic HiSGRA's have been constructed in higher dimensions~\cite{Vasiliev:2003ev,Sagnotti:2005ns,Sharapov:2019pdu,Sharapov:2019vyd} within the formal deformation scheme.  Their supersymmetric versions are known at the linearized level in  $5D$ ~\cite{Sezgin:2001yf} and $7D$~\cite{Sezgin:2002rt}, but their fully nonlinear version remain to be constructed. 

One of the notorious problems in string theory is the difficulty in obtaining a de Sitter ($dS$) solution.  An interesting mechanism was proposed long ago by KKLT \cite{Kachru:2003aw} but this has encountered serious obstacles, the most recent one noted in \cite{Lust:2022lfc}.  In HiSGRA, however,  both signs of the
cosmological constant are allowed, and they can be chosen to be very small. This fact has been exploited in~\cite{Anninos:2011ui} where it was conjectured that 
HiSGRA in $dS_4$ is holographically dual to a $CFT_3$ living on the spacelike boundary of $dS_4$ at future timelike infinity. Further progress has been made in~\cite{Das:2012dt,Anninos:2013rza,Anninos:2017eib,David:2020ptn}. This line of development clearly deserves further studies. 

Ultimately, a (quasi-)local Effective Field Theory (EFT) that provides an $S$ matrix as the well-defined observable in flat spacetime, is needed.  On the other hand, there is no known fully nonlinear higher spin extension of Einstein's gravity in 4D that admits flat spacetime as a vacuum solution.  Once again the importance of spontaneous breaking of higher spin symmetry enters the picture to achieve the stated goal.  Another avenue to pursue may be the utilization of the domain-wall solutions of a higher dimensional HiSGRA. While higher spin  field equations seem to be prohibitively  difficult to solve, a remarkable framework exists in which exact solutions have been found, and they contain domain-wall solutions among others (see~\cite{Iazeolla:2017dxc} for a review). Their in-depth study  is in progress and much remains to be done in this direction.

\subsection{Higher Spin Symmetry for Conformal Field Theory}
As it was already reviewed, a number of encouraging results, see e.g.~\cite{Maldacena:2012sf,Alday:2015ota,Skvortsov:2015pea,Giombi:2016hkj,Giombi:2016zwa,Alday:2016jfr,Giombi:2017rhm,Skvortsov:2018uru,Turiaci:2018nua,Sharapov:2018kjz,Li:2019twz,Binder:2021cif,Jain:2020puw,Gerasimenko:2021sxj}, gives a strong support to the idea of the slightly-broken higher spin symmetry being realized in Chern-Simons matter theories at least in the large-$N$ limit. This immediately leads to two challenges. (1) what slightly-broken higher spin symmetry means as a symmetry? It has been mostly explored via the study of the non-conservation equation for higher spin currents, where the crucial observation \cite{Maldacena:2012sf} is that the conservation can be broken only by double- and triple-trace operators that are built of the higher spin currents themselves. The latter allows one to apply the non-conservation law inside correlation functions deriving, thereby, strong constraints for the correlators with the help of the large-$N$ factorization. Clearly, slightly-broken higher spin symmetry is a new type of an infinite-dimensional symmetry that extends the conformal symmetry at least in three dimensions. It is tempting to argue that in the context of vector models it can be somewhat analogous to Virasoro symmetry. (2) it is critically important to extend the applications of the slightly-broken higher spin symmetry to higher point correlation functions  to constrain them and obtain explicit expressions. Since the constraints imposed by the slightly-broken higher spin symmetry are insensitive to whether the higher spin currents are built of bosonic or fermionic matter, the three-dimensional bosonization can be proved without having to resort to any concrete microscopical description in terms of Chern-Simons matter theories. If the latter is achieved, the next goal would be to extend the slightly-broken higher spin symmetry beyond the large-$N$ limit. It looks encouraging that the non-conservation equation is also true for small values of $N$, while the higher spin currents have (numerically) small anomalous dimensions even for the $N=1$ Ising model.  

In a similar vein, the higher spin symmetry of the weakly coupled $\mathcal{N}=4$ SYM theory can be explored, see \cite{Bianchi:2003wx,Beisert:2004di} for earlier works. It does not, however, follow the slightly-broken higher spin symmetry idea, since the conservation of the $\mathcal{N}=4$ SYM higher spin currents is not broken by the double-trace operators built of the higher spin currents themselves. The spectrum of operators also contain infinitely many operators on top of the currents. The latter can be organized into certain multiplets of the higher spin algebra. Understanding the mechanism of the higher spin breaking in $\mathcal{N}=4$ SYM can lead to new (as compared to integrability \cite{Beisert:2010jr}) computational techniques to get anomalous dimensions at least at weak coupling and can shed more light on the tensionless limit of string theory \cite{Gaberdiel:2021jrv,Gaberdiel:2021qbb}.

\subsection{Higher Spin Techniques for Black Hole Scattering}
\label{bh}
The recent experimental discovery of gravitational waves by LIGO \cite{LIGOScientific:2016aoc} and the rapid development of gravitational-wave physics leads to new challenges for theoretical physics. 
Precision computations of gravitational-wave templates are necessary for interpreting data from LIGO \cite{LIGOScientific:2014pky}, VIRGO \cite{VIRGO:2014yos} and similar experiments. While numerical relativity is necessary to describe the final stages of a merger, this approach is computationally expensive. Hence, there is a high demand for efficient analytic methods, particularly for the so-called inspiral phase of the merger in which the two compact object emitting gravitational waves (i.e., black holes or neutron stars) are still well separated. It should be noted that, in the mergers that lead to gravitational waves observed by LIGO and similar experiments, the inspiral phase accounts by far for the largest portion of the observed signal. 
    
    Among the various analytic approaches that have been developed to carry out calculations relevant to gravitational-wave physics, the ones most closely related to HiSGRA draw from  the EFT framework. These approaches describe well-separated black holes  with an effective worldline theory of point masses  coupled to gravity (see Refs. \cite{Goldberger:2004jt,Porto:2005ac, Neill:2013wsa} for their foundations). 
    While worldline EFT techniques give results in a post-Newtonian expansion (i.e., weak fields, non-relativistic velocities), more recently, methods based on scattering amplitudes have also been developed.  These approaches build on techniques developed for scattering amplitude calculations in relativistic quantum field theories and give results which are naturally organized in a post-Minkowskian expansion (i.e.\ an expansion in Newton's constant).
    A full account of the application of amplitude methods to calculations related to gravitational-wave physics is beyond the scope of this paper; the interested reader should refer to the white paper \cite{Buonanno:2022pgc} 
    (see also Ref. \cite{Adamo:2022dcm} for an overview of modern methods for amplitudes in gravity).

    The standard route to obtain the effective Hamiltonian of a binary system requires matching the classical limit of a QFT amplitude with the amplitude calculated from an EFT with free parameters (see, e.g., Ref.~\cite{Cheung:2018wkq,Bern:2019nnu,Bern:2019crd}). This matching calculation allows to fix the desired two-body effective Hamiltonian, which can then be used to describe the bound system and obtain the desired gravitational-wave template. Alternative procedures have also been formulated that allow to obtain data for bound orbits by analytic continuation
    \cite{Kalin:2019rwq}.
    However, the field theory employed to calculate amplitudes in the classical limit needs to be chosen carefully. In case of mergers in which the black holes do not carry spin, a theory with massive scalars is sufficient---indeed many of the results from amplitude methods are obtained in this case, including results up to ${\cal O}(G^4)$ \cite{Bern:2021dqo,Bern:2021yeh}.  
    
    An important open problem is the inclusion of spin effects in the amplitude framework (see
    also Ref. \cite{Levi:2018nxp} for a comprehensive review of spin effects in the post-Newtonian expansion using worldline EFTs and Refs. \cite{Levi:2019kgk,Levi:2020kvb,Levi:2020uwu,Mogull:2020sak,Levi:2020lfn,Jakobsen:2021lvp,Kim:2021rfj,Jakobsen:2021zvh,Jakobsen:2022fcj,Edison:2022cdu} for recent results using worldine approaches). Some results for spin corrections to the two-body effective potential for a binary system were carried out in Refs. \cite{Chung:2018kqs,Guevara:2018wpp,Guevara:2019fsj,Maybee:2019jus,Chung:2020rrz,Bern:2020buy,Guevara:2020xjx,Kosmopoulos:2021zoq,Chen:2021qkk,Aoude:2021oqj,Bern:2022kto,Aoude:2022trd}. One of the major challenges that have emerged is however identifying the correct QFT of massive spinning fields that needs to be used to compute the correct scattering amplitudes. Scattering amplitudes corresponding to a spinning (Kerr) black hole emitting a graviton are known at three points \cite{Arkani-Hamed:2017jhn,Arkani-Hamed:2019ymq,Guevara:2020xjx}. However, higher point (Compton) amplitudes between massive fields of generic spin and two or more gravitons are necessary for precision calculations, and these amplitudes are affected by a contact-term ambiguity. Attempts to resolve this ambiguity were carried out in Refs. \cite{Chiodaroli:2021eug,Falkowski:2020aso,Aoude:2021oqj}, but research in this direction is ongoing. This is precisely where the HiSGRA framework is necessary---identifying from physical principles the correct action to model the coupling of spinning black-holes to gravity would provide the building-blocks necessary for applying amplitude methods to problems in gravitational radiation. 
    By providing a gauge-invariant
description of massive higher spin particles, HiSGRA gives systematic tools for tackling this problem  \cite{Zinoviev:2001dt,Zinoviev:2006im,Zinoviev:2008ck}.
The fact that HiSGRA might be applied to solve challenging problems in amplitudes and gravitational radiation is indicative of its close connection with other subfields in theoretical and gravitational physics and underlines the potential for mutually-beneficial synergies.

\subsection{Non-relativistic Higher Spin Symmetry}

The quantum many-body problem of a non-relativistic two-component Fermi gas with short-range attractive interactions is a longstanding problem in condensed matter physics. At low temperature, the system is known to be superfluid and undergoes a smooth crossover from the Bardeen-Cooper-Schrieffer (BCS) to the Bose-Einstein-Condensate (BEC) regime as the two-body attraction is increased \cite{Bloch:2008zzb,Giorgini:2008zz}. In particular, there is a specific regime in between BCS and BEC, known as the ``unitary Fermi gas'' of special theoretical interest because, on the one hand, it is strongly coupled and no obvious small parameter is available precluding the reliable application of a perturbative expansion but, on the other hand, a characteristic of the unitary Fermi gas in vacuum is its invariance under the Schr\"odinger group of \cite{Niederer:1972zz,Hagen:1972pd}, which extends the Galilean group of non-relativistic symmetries by scale transformations and expansions (a non-relativistic analogue of special conformal transformations). This non-relativistic conformal symmetry of the unitary Fermi gas allowed \cite{Son:2008ye,Balasubramanian:2008dm} to apply the methods of gauge-gravity duality to this system. While these seminal papers triggered an intensive search for the holographic duals of various non-relativistic systems originating from condensed matter physics, a holographic description of the unitary fermions still remains tantalising. In \cite{Bekaert:2011cu}, inspired by the method of null dimensional reduction \cite{Duval:1984cj,Duval:2009vt} and the conjectured AdS dual of the free/critical $O(N)$ model, a holographic dual description of the ideal/unitary Fermi gas was attempted. However, the corresponding bulk gravity theory remains elusive; it deserves further investigations and possible improvements.

Another route to apply gauge-gravity techniques to condensed matter systems was pioneered in \cite{Bagchi:2009my}: the key idea is to study systems that display a different conformal-like extension of the Galilei algebra that, differently from the Schr\"odinger algebra, can be obtained from the contraction of the relativistic conformal algebra. Recently, a higher spin extension of the conformal Galilei algebra was proposed \cite{Campoleoni:2021blr} and this may pave the way to the holographic description of other condensed-matter systems with infinite-dimensional symmetries, along the lines of the quantum Hall effect \cite{Cappelli:1992yv}.

More generally, there is a number of condensed matter systems, 
e.g.\ massive Chern-Simons matter theories, fractons \cite{Pretko:2020cko}, 
Kondo systems, unitary Fermi gas and quantum Hall effect, 
where it could be possible to either directly exploit (possibly, slightly-broken) 
higher spin symmetry or to apply many higher spin techniques developed to work 
with tensorial degrees of freedom. 
In fact, in \cite{Boulanger:2013naa,Boulanger:2015uha} several
2+1 dimensional Chern--Simons models were constructed that describe in a unified way 
fractional-spin fields coupled to tensorial higher-spin gravity and 
internal non-abelian gauge fields. The second quantisation of fractional-spin fields
leads to anyons, therefore the results of \cite{Boulanger:2013naa,Boulanger:2015uha}
suggest the possibility of a dual HiSGRA description of anyons. 
Finally, thanks to the remarkable simplifications displayed by higher-spin theories 
in three dimensions, various non-relativistic holographic setups have been  
proposed in that context \cite{Gary:2012ms,Afshar:2012nk,Gary:2014mca,Chernyavsky:2019hyp}.

\subsection{Higher Spin Gravity and Mathematics}

\paragraph{Deformation quantization, non-commutative geometry.}
Higher spin symmetries have long been known to result from the canonical Deformation Quantization of Poisson manifolds or alternatively (and equivalently), as algebras of higher symmetries of free field equations of motion~\cite{Nikitin1991,Eastwood:2002su,Bekaert:2009fg,Michel2014}. However, gauging higher spin symmetry leads to new twists that correspond to an extensions of Deformation Quantization --- Deformation Quantization of Poisson Orbifolds, which is still an open problem in mathematics. It is not even clear what would be the proper analog of Kontsevich formality for Poisson Orbifolds \cite{Tiang,Halbout}. Nevertheless, some simple cases of Weyl algebras extended with various groups of symplectomorphisms were treated by different methods \cite{Wigner1950,Yang:1951pyq, Boulware1963, Gruber, Mukunda:1980fv,Pope:1990kc,Bieliavsky:2008mv,Joung:2014qya,Korybut:2014jza,Basile:2016goq,Sharapov:2017lxr,Sharapov:2018hnl,Korybut:2020vmm}. The latter provides a great number of examples that are both useful for HiSGRA applications and can lead to the development of new techniques for deformation quantization. 

\paragraph{Geometry of  Gauge Theories.}
From its early days HiSGRA was challenging the mathematical foundations of QFT and now continues to do so. A characteristic feature of HiSGRA models is the intricate interplay between the spacetime and field-space geometry, rigid and gauge symmetries, and somewhat generalised notion of locality and variational principle. All these call for developing an adequate quantization formalism unifying the existing BV-BRST-like approaches~\cite{Barnich:2010sw,Grigoriev:2010ic,Cattaneo:2012qu,Alkalaev:2013hta,Cattaneo:2015vsa,Mnev:2019ejh}, including AKSZ sigma models, and frame-like formulations and extending them beyond the standard jet-bundle setting and covering local gauge theories on manifolds with (asymptotic) boundaries. Interesting results in this direction have been already obtained in~\cite{Barnich:2010sw,Grigoriev:2010ic,Cattaneo:2012qu,Bekaert:2013zya,Cattaneo:2015vsa,Sharapov:2016sgx,Mnev:2019ejh}. 

A proper BV-BRST-like description of background fields and background independence \cite{Horowitz:1986dta,Grigoriev:2006tt,Dai:2008bh,Bekaert:2013zya,Adamo:2014wea,Bonezzi:2018box,Bonezzi:2010jr,Grigoriev:2021bes} is expected to uncover the still somewhat elusive higher spin geometry~\cite{Sezgin:2011hq,Boulanger:2015kfa}. These developments are also expected to shed some light on the relation of HiSGRA to string (field) theory and its background independence.

Applications to HiSGRA have also resurrected the interest in the inverse problem of variational calculus, especially in the context of gauge systems~\cite{Khavkine2012,Alkalaev:2013hta,Boulanger:2012bj,Sharapov:2016qne,Sharapov:2016sgx,Grigoriev:2021wgw} and has triggered the development of the presymplectic AKSZ approach~\cite{Alkalaev:2013hta,Grigoriev:2016wmk,Grigoriev:2020xec} which turned out remarkably fruitful~\cite{Sharapov:2021drr} in the HiSGRA context. Promising future directions include developing efficient quantization scheme for presymplectic AKSZ systems.

\paragraph{AKSZ quantization of Formal HiSGRA.}

The BV-BRST extension of the formal HiSGRA's (most notably of the original Vasiliev's equations) naturally has the form of an AKSZ sigma model~\cite{Barnich:2005ru,Boulanger:2012bj,Alkalaev:2014nsa,Sharapov:2021drr}. While Fronsdal's formulation of higher spin gravity is based on a perturbative expansion around AdS with observables given by Witten diagrams, it was shown in \cite{Boulanger:2015ova}, under natural assumptions, that Vasiliev's classical field equations, which provide a fully nonlinear background independent formulation of HiSGRA, does not lend itself to computation of such diagrams.
Instead, thinking of the moduli space of this version of higher spin theory as 
deformations of fibered noncommutative background geometries of first-quantized 
models \cite{Arias:2016agc} --- which rhymes well with \cite{Sharapov:2021drr} ---  
leads to second-quantized AKSZ sigma models \cite{Boulanger:2015kfa,Bonezzi:2016ttk} 
formulated directly in terms of the higher spin master fields, interpreted as horizontal 
forms.
It has been proposed that their effective actions have perturbative expansions in terms of 
zero-form charges \cite{Sezgin:2005hf} reproducing the first-quantized amplitudes 
\cite{Engquist:2005yt} known to give rise to the desired holographic higher spin amplitudes 
once evaluated on states in appropriate representations \cite{Giombi:2010vg,Colombo:2010fu,Colombo:2012jx,Didenko:2012tv,Didenko:2013bj,Bonezzi:2017vha}. 
The completion of this perturbative scheme already at tree-level remains an interesting open problem.

\paragraph{Conformal (higher spin) Geometry.}
Conformal geometry is one of oldest branches of mathematics, yet it is still full of open problems that are easy to formulate. The most basic questions are about classification of conformal invariants and, closely related to them, conformal anomalies and conformally invariant differential operators. Even though most of the physics is not conformally invariant (in the sense of conformal geometry, i.e., with respect to local Weyl transformations), conformal geometry has surprisingly many connections to theoretical physics. 

One existing class of HiSGRA, conformal HiSGRA \cite{Segal:2002gd,Tseytlin:2002gz,Bekaert:2010ky}, opens up new ways of approaching these old problems. By its very definition, each conformal HiSGRA, being a higher spin extension of conformal gravity, leads to a variety of conformally-invariant operators~\cite{Nutma:2014pua,Grigoriev:2016bzl,Beccaria:2017nco,Kuzenko:2018lru,Kuzenko:2019ill,Kuzenko:2019eni,Kuzenko:2020jie}. 
Conformal HiSGRA also gives a new perspective on conformal geometry since the conformal symmetries become a subalgebra of higher spin transformations. Moreover, the HiSGRA naturally leads to a higher spin extension of (conformal) geometry~\cite{Segal:2002gd,Bekaert:2010ky,Grigoriev:2016bzl,Bekaert:2017bpy,Grigoriev:2018wrx}.

\paragraph{Twistor theory.}
Twistor theory is well adapted to self-dual backgrounds and theories. It had also been somewhat ahead of the main higher spin developments by offering a number of important results that are valid for fields of any spin 
\cite{Penrose:1965am,Hughston:1979tq,Eastwood:1981jy,Woodhouse:1985id,Hitchin:1980hp}. The recently constructed self-dual higher spin theories \cite{Ponomarev:2016lrm,Ponomarev:2017nrr,Hahnel:2016ihf,Adamo:2016ple} together with higher spin extensions of self-dual Yang-Mills and self-dual gravity theories \cite{Krasnov:2021nsq,Tran:2021ukl} point towards twistor methods being relevant and set new challenges for twistor theory. For example, Chiral higher spin gravity is a generalisation of both of self-dual Yang-Mills and self-dual Gravity theories and, in fact, it shares many features with them. One crucial property of SDYM and SDGR is that they are integrable systems, a property made manifest through their reformulation in twistor space --- this is the content of Ward's Theorem for SDYM \cite{Ward:1977ta} and Penrose non-linear graviton theorem for SDGR \cite{Penrose:1976js}. What is more, twistor variational principles for these theories have been useful to derive features of the corresponding scattering amplitudes that are otherwise inaccessible (see, e.g., reviews \cite{Adamo:2011pv,Atiyah:2017erd}). This begs the questions: could it be that Chiral HiSGRA (and its non-chiral completion) admit covariant variational principles in twistor space (possibly without any spacetime equivalent)? If so, what could this geometrical reformulation in twistor space tell us about HiSGRA? The same questions apply to the four-dimensional conformal HiSGRA and its self-dual truncations \cite{Hahnel:2016ihf,Adamo:2016ple}.

\section{Outlook}

In this brief review, we have provided a partial list of results achieved so far in higher spin theory, exciting recent progress, and selected topics for future directions. Higher spin theories are expected to have impact on how we treat quantum gravity, spacetime and black holes, in a way that differs profoundly from the treatment of Einstein's gravity. In particular, higher spin symmetries, non-localities and non-commutative extension of spacetime play key roles in their description, and several mathematical developments that emerge in the study of higher spin theories have the potential of boosting the interface with the mathematics discipline. 

\section*{Acknowledgments}
\label{sec:Aknowledgements}

We are grateful to Yannick Herfray and Per Sundell for valuable contributions to this review. 
A.C.\ and E.S.\ are Research Associates of the Fund for Scientific Research (FNRS), Belgium. The work of N.B. was partially supported by FNRS under Grant No.\ T.0022.19. The work of A.C.\ was partially supported by FNRS under Grant No.\ F.4503.20. The work of E.S. was partially supported by the European Research Council (ERC) under the European Union’s Horizon 2020 research and innovation programme (grant agreement No 101002551) and by FNRS under Grant No.\ F.4544.21. The work of E.Se.\ is supported in part by the NSF grants PHY-1803875 and PHYS-2112859.
The work of M.C.\ is supported by the Swedish Research Council under grant 2019-05283.\\

\footnotesize
\providecommand{\href}[2]{#2}\begingroup\raggedright\endgroup

\end{document}